%% file: prl.tex
\documentclass[aps,prl,twocolumn,showpacs,groupedaddress]{revtex4}  
\usepackage{graphicx}  
\usepackage{dcolumn}   
\usepackage{bm}        
\usepackage{amssymb}   

\def\ttbar{t\overline{t}}
\def\bbbar{b\overline{b}}
\def\ppbar{p\overline{p}}

\def\pythia{{\sc pythia}}
\def\alpgen{{\sc alpgen}}
\def\tauola{{\sc tauola}}
\def\evtgen{{\sc evtgen}}
\def\geant{{\sc geant}}
\def\feynhiggs{{\sc feynhiggs}}

\def\mcfm{{\sc mcfm}}
\newcommand{\met}       {\mbox{$\not\!\!E_T$}}
\def\hto{\hphantom{000}}

\begin{document}


\hspace{5.2in} \mbox{FERMILAB-PUB-09-612-E}

\title{Search for the associated production of a \boldmath{$b$} quark and a neutral supersymmetric Higgs
  boson which decays to tau pairs}

\input list_of_authors_r2.tex  
\date{December 4, 2009}

\begin{abstract}
We report results from a search for production of a neutral
Higgs boson in association with a $b$ quark. We search for Higgs decays to $\tau$ 
pairs with one $\tau$ subsequently decaying to a muon and the other to
hadrons.  The data correspond to 2.7~fb$^{-1}$ of $\ppbar$ collisions recorded by the
D0 detector at $\sqrt{s} = 1.96$~TeV.  The data are found to be consistent
with background predictions.  The result allows us to exclude a significant region of 
parameter space of the minimal supersymmetric model. 
\end{abstract}

\maketitle 

The current model of physics at high energies, the standard model (SM), has
withstood increasingly precise experimental tests, although the Higgs boson
needed to mediate the breaking of electroweak symmetry has not been found.
Despite the success of the SM, it has several shortcomings.
Theories invoking a new fermion-boson symmetry, called 
supersymmetry~\cite{b-susy} (SUSY),
provide an attractive means to address some of these including the hierarchy
problem and nonunification of couplings at high energy.  In
addition to new SUSY-specific partners to SM particles, these theories have an extended
Higgs sector.  In the minimal supersymmetric standard model (MSSM) there are two Higgs doublet fields
which result in five physical Higgs bosons: two neutral
scalars ($h, H$), a neutral pseudoscalar ($A$) and two charged Higgs
bosons ($H^\pm$).  The mass spectrum of the Higgs bosons is determined
at tree level by two parameters, typically chosen to be $\tan\beta$, the
ratio of the vacuum expectation values of up-type and down-type scalar fields and
$M_A$, the mass of the physical pseudoscalar. Higher order corrections are
dominated by the Higgsino mass
parameter $\mu$ and the mixing of scalar top quarks.

In this Letter, we present a search for neutral Higgs bosons (collectively
denoted $\phi$) produced in association with a $b$ quark.  The specific Higgs 
boson decay mode used in this search is $\phi\to\tau\tau$ with one of the $\tau$ leptons
subsequently decaying via $\tau\to\mu\nu_{\tau}\nu_{\mu}$ (denoted $\tau_\mu$) and the second
via $\tau\to\ \mathrm{hadrons}+\nu_\tau$ (denoted $\tau_{h}$).  
In the MSSM the Higgs coupling to down-type fermions is enhanced by a factor $\propto \tan\beta$ 
and thus the Higgs production cross section is enhanced by a factor $\propto\tan^2\beta$ relative to the
SM, giving
potentially detectable rates at the Tevatron.  Two of the three neutral Higgs bosons have nearly
degenerate masses over much of the parameter space, effectively giving another factor of
two in production rate.  
A previous search in this final state was carried out by the D0
experiment~\cite{b-p14}.  Searches in the complementary channels
$\phi Z/\phi\phi\to b\bar{b}\tau\tau,\tau\tau b\bar{b}$~\cite{b-LEP2},
$\phi\to\tau\tau$~\cite{b-tautau,b-tautau-cdf}, and $\phi b\to
\bbbar b$~\cite{b-hb-bbb,b-hb-bbb-cdf} have also been carried out by the LEP, D0, and
CDF experiments.  By searching in complementary channels we reduce overall sensitivity to the particular
details of the model.   The $b\tau\tau$ final state is less sensitive to SUSY radiative
corrections than the $\bbbar b$ final state, and has greater sensitivity at 
low Higgs mass than the $\phi\to\tau\tau$ channel, as the $b$-jet in the final state 
reduces the $Z\to\tau\tau$ background.  Furthermore, an additional complementary channel will contribute 
to an even stronger exclusion when combining different searches.   
The result presented in this Letter uses an
integrated luminosity of 2.7~fb$^{-1}$ which is eight times that
used for the
previous result in this channel.  Because of analysis improvements,
the gain in sensitivity compared to the prior result is
greater than expected from the increased integrated luminosity only.  We also extend
the Higgs mass search range relative to the previous result in this channel. 

The D0 detector~\cite{d0det} is a general purpose detector located at Fermilab's Tevatron
$\ppbar$ collider.  The Tevatron operates at a center of mass energy of 1.96 TeV.
This analysis relies on all aspects of the detector: tracking,
calorimetry, muon detection, the ability to identify detached vertices and the
luminosity measurement.  

This search requires reconstruction of muons, hadronic $\tau$ decays, jets
(arising from $b$ quarks) and neutrinos.  Muons are identified using track segments
in the muon system and are required to have a track reconstructed in the 
inner tracking system which is
close to the muon-system track segment in $\eta$ and $\varphi$.  Here $\eta$ is the
pseudorapidity and $\varphi$ is the azimuthal angle in the plane perpendicular
to the beam.  Jets are reconstructed from calorimeter information using the D0 Run II cone algorithm~\cite{b-jetalg} 
with a radius of $R=0.5$ in $(y,\varphi)$ space, where $R=\sqrt{(\Delta y^2+\Delta \varphi^2)}$ and $y$ is the rapidity.  
Jets are additionally identified as being consistent with decay of a $b$-flavored hadron ($b$-tagged) if the 
tracks aligned with the calorimeter jet have high impact parameter or form a vertex separated from the primary
interaction point in the plane transverse to the beam as determined by a neural
network (NN$_b$) algorithm~\cite{b-btag}.  Hadronic $\tau$ decays are
identified~\cite{b-ztautau} as clusters of energy in the calorimeter reconstructed~\cite{b-jetalg} using
a cone algorithm of radius $R=0.3$ which have associated
tracks.  The $\tau$ candidates are then categorized as being one of three types
which correspond roughly to one-prong $\tau$ decay with no $\pi^0$s (called Type
1), one-prong decay with $\pi^0$s (Type 2) and multiprong decay (Type
3).  A final identification requirement is based on the output value of a
neural network (NN$_\tau$) designed to separate $\tau$ leptons from jets.  The missing
transverse energy $\met$ is used to infer the presence of neutrinos.  The
$\met$ is the negative of the vector sum of the transverse energy of calorimeter cells 
satisfying $|\eta|<3.2$. $\met$ is corrected for the energy scales of reconstructed final state objects, including muons.

Signal acceptance and efficiency are modeled using simulated SM $\phi b$ events generated with the \pythia\ event generator~\cite{b-pythia}
requiring the $b$ quark to satisfy $p_T>15$~GeV/$c$ and $|\eta|<2.5$ and using the 
CTEQ6L1~\cite{b-cteq} parton distribution functions (PDF).  The 
\tauola~\cite{b-tauola} program is used to model $\tau$ decay and
\evtgen~\cite{b-evtgen} is used to decay $b$ hadrons.  The dependence of the
Higgs boson decay width on $\tan\beta$ is included by reweighting \pythia\
samples, and the kinematic properties are reweighted to the prediction of
the NLO program \mcfm~\cite{b-mcfm}.  The generator outputs 
are passed through a detailed detector simulation based on
\geant~\cite{b-geant}.  Each \geant\ event is combined with collider
data events recorded during a random beam crossing to model the effects of detector noise, 
pileup, and additional $p \bar p$ interactions.  The
combined output is then passed to the D0 event reconstruction program.
Simulated signal samples are generated for different Higgs masses ranging from 90 to
320~GeV/$c^2$. 

Backgrounds to this search are dominated by $Z$+jets, $\ttbar$,
and multijet (MJ) production.  In the MJ background the apparent leptons primarily come
from semileptonic $b$ hadron decays, not $\tau$ decays. 
Additional
backgrounds include $W$+jets events,
SM diboson production and single top quark production.  Except for
the MJ contribution, all background yields are estimated using simulated
events, with the same processing chain used for signal
events.  The $Z$+jets, $W$+jets and $\ttbar$ samples are generated using
\alpgen~\cite{b-alpgen} with \pythia\ used for fragmentation.  The diboson 
samples are generated using \pythia.  For simulated samples in which there is only one 
lepton arising from the decay of a $W$ boson or from 
$\ttbar\to\ell+$jets, the second lepton is either a jet misidentified as a $\tau$
or a muon+jet system from heavy flavor decay in which the muon is misidentified
as being isolated from other activity.  

Corrections accounting for differences between data and
the simulation are applied to the simulated events.  The corrections are derived from control data
samples and applied to object identification efficiencies, trigger
efficiencies, primary $\ppbar$ interaction position (primary vertex) and the transverse momentum 
spectrum of $Z$ bosons.  
After applying all corrections, the yields for signal
and each background are calculated as the product of the acceptance times
efficiency determined from simulation, luminosity and predicted cross sections.

The initial analysis step is selection of events recorded by at
least one trigger from a set of single muon triggers for data taken
before the summer of 2006.  For data taken after summer 2006 we require at least one trigger 
from a set of single muon triggers and muon plus hadronic $\tau$ triggers.  The 
average trigger efficiency for signal events is approximately 65\%\ for both data epochs.

After making the trigger requirements a background-dominated pre-tag sample is
selected by requiring 
a reconstructed primary vertex
for the event with at least three tracks,
exactly one reconstructed hadronic $\tau$, exactly one isolated muon,
and at least one jet.  This analysis requires the $\tau$ candidates to satisfy 
 $E_T^\tau>10$~GeV, $p_T^\tau>7(5)$~GeV/$c$ and $NN_\tau>0.9$ for Type 1(2) taus, $E_T^\tau>15$~GeV, 
$p_T^\tau>10$~GeV/$c$ and $NN_\tau>0.95$ for Type 3 taus.  Here $E_T^{\tau}$ is the transverse energy 
of the $\tau$ measured
in the calorimeter, $p_T^\tau$ is the transverse momentum sum of the associated track(s).
The muon must satisfy $p_T^\mu>12$~GeV/$c$ and
$|\eta|<2.0$.
It is also required to be isolated from
activity in the tracker and calorimeter~\cite{b-hpp}.
Selected jets
have $E_T>15$~GeV, $|\eta|<2.5$.  The $\tau$, the muon and jets must all be
consistent with arising from the same primary vertex and be separated from each
other by $\Delta R > 0.5$.  In addition, the muon and $\tau$ are required to have
opposite charge, and the $(\mu,\met)$ mass variable $M \equiv \sqrt{2\met
E_\mu^2/p_T^\mu(1-\cos(\Delta\varphi(\mu,\met)))}$ must satisfy $M < 80,\ 80,\ \mathrm{and}\
60$~GeV$/c^{2}$ for events with $\tau$s of Type 1, 2 and 3 respectively.  Here $E_\mu$ is
the energy of the muon,
and
$\Delta\varphi$ is the opening angle between the $\met$ and muon in the plane
transverse to the beam direction.

A more restrictive $b$-tag subsample with improved signal to background ratio is 
defined by demanding that at least one jet in each event is consistent with
$b$ quark production~\cite{b-btag}.  The $b$-jet identification efficiency in signal events is about 35\%\ and the
probability to misidentify a light jet as a $b$ jet is 0.5\%.  

All backgrounds except MJ are derived from simulated events as described
earlier.  The MJ background is derived from control data samples.  A parent
MJ-enriched control sample is created by requiring a muon, $\tau$, and jet as
above, but with the muon isolation requirement removed and with a lower quality
($0.3\leq NN_{\tau}
\le 0.9$) $\tau$ selected.  This is then used to create a $b$-tag subsample
which requires at least one of the jets to be identified as
a $b$ jet with the same $b$ jet selection as earlier.  The
residual contributions from SM backgrounds are subtracted from the MJ control samples using simulated events.

To determine the MJ contribution in the pre-tag analysis sample, a data 
sample is used that has the same selection as the pre-tag analysis sample except that the 
muon and $\tau$ charges have the 
same sign.  This same-sign (SS) sample is dominated by MJ events.  After making a 
subtraction of other SM background processes which contribute to
this sample, the number of MJ events in the opposite-sign (OS) signal region is
computed by multiplying the SS sample by the OS/SS ratio, $1.05\pm0.02$, determined in a 
control sample selected by requiring a non-isolated muon..

For the $b$-tag analysis sample, statistical limitations require
a different approach for the MJ background evaluation than for the pre-tag sample.  For the
$b$-tag sample, two methods are used.  For the first method, the per jet probability
$P_{tag}$ that a jet in the SS MJ control subsample would be identified as a $b$ jet is 
determined as a  function of jet $p_T$.
We apply $P_{tag}$ to the jets in the SS pre-tag sample to 
determine the yield in the $b$-tag sample.  
For the second method, the MJ background is determined by multiplying the
$b$-tag MJ control sample yield by two 
factors: (1) the probability that the non-isolated muon would be identified as 
isolated, and (2) the ratio of events with a $\tau$ candidate passing the $NN_{\tau}$ requirements to events with $\tau$ 
candidates having $0.3\leq NN_{\tau} < 0.9$ as determined in 
a separate control sample.
The final MJ contribution in the $b$-tag analysis sample is determined using the MJ shape
from the first method with the normalization equal to the average of the two methods.  
We include the normalization difference between the two
methods in the systematic uncertainty on the MJ contribution.

The signal to background ratio is further improved using multivariate techniques.
Two separate methods are used, one to address the $\ttbar$ background and one
to reduce the MJ background.  For the $\ttbar$ background, a neural network
($\mbox{\em NN}_{top}$) is constructed using $H_T \equiv \Sigma_{jets}E_T$, $E_{tot} \equiv
\Sigma_{jets} E + E_\tau + E_\mu$, the number of jets and
$\Delta\varphi(\mu,\tau)$ as inputs.  For the MJ background, a simple joint
likelihood discriminant ($\mbox{\em LL}_{MJ}$) is constructed using $p_T^\mu$, $p_T^\tau$, $\Delta
R(\mu,\tau)$, $M_{\mu\tau}$ and $M_{\mu\tau\nu}$.  Here $M_{\mu\tau}$ denotes the
invariant mass of the muon and tau, and $M_{\mu\tau\nu}$ is the invariant mass 
computed from the muon, $\tau$, and $\met$ momentum vectors.
The final analysis sample is defined by selecting rectangular regions in the $\mbox{\em NN}_{top}$
versus $\mbox{\em LL}_{MJ}$ plane.  The regions have been identified 
for each $\tau$ type and each Higgs boson mass point separately by optimizing the search sensitivity 
using simulated events.  The
signal to background ratio improves by up to a factor of two when applying these requirements.


Table~\ref{t-yields} shows the predicted background and observed data yields in the analysis samples.  
Between 5\% and 10\% of $\phi\to\tau_\mu \tau_h$ decays are selected depending on $M_\phi$.
\begin{table}
\begin{center}
  \begin{tabular}{l|ccc}
           & \hto Pre-tag\hto & \hto $b$-tagged \hto & \hto Final \hto \\ \hline
 $\ttbar$  & $66.0\pm1.3$       & $39.6\pm0.8$         & $20.3 \pm0.6$ \\
 Multijet  & $549\pm26$       & $38.5\pm2.3$         & $28.1\pm1.9$ \\
 $Z(\to\tau\tau)+\mathrm{jets}$   & $1241\pm8$       & $18.8\pm0.3$         & $16.3\pm0.3$ \\\hline
 Other Bkg & $ 267\pm6$       & $ 5.1\pm0.1$         & $ 4.1\pm0.1$ \\\hline
 Total Bkg & $2123\pm28$      & $102\pm2.4$          & $68.8\pm2.0$ \\ \hline\hline
 Data      & 2077             &  112                 & 79      \\ \hline\hline
 Signal    & $14.4\pm0.3$     & $4.8\pm0.1$         & $4.6\pm 0.1$ \\
  \end{tabular}
  \caption{Predicted background yield, observed data yield and predicted signal
    yield and their statistical uncertainties at three stages of the analysis.
    The signal yields are calculated
    assuming $\tan\beta = 40$ and a Higgs mass of 120 GeV/$c^{2}$ for the
    $m_h^{max}$ and $\mu = -200$ GeV$/c^{2}$ scenario.\label{t-yields}}
\end{center}
\end{table}

Systematic uncertainties arise from a variety of sources.  Most are evaluated
using comparisons between data control samples and predictions from
simulation. The uncertainties are divided into two categories: (1) those which
affect only normalization, and (2) those which also affect the shape of
distributions.  The sources in the first category include the  luminosity (6.1\%),
muon identification efficiency (4.5\%), $\tau_{h}$ identification (5\%, 4\%, 8\%),
$\tau_{h}$ energy calibration (3\%),  the $\ttbar$ and single top
cross sections (11\% and 12\%), diboson cross sections (6\%), $Z$+($u$,$d$,$s$,$c$)
rate (+2\%, -5\%) and the $W+b$ and $Z+b$ cross sections (30\%); those in the second include 
jet energy calibration (2\%-4\%),
$b$-tagging (3\%-5\%), trigger (3\%-5\%), and MJ background (33\%, 12\%, 11\%).  For sources
with three values, the values correspond to $\tau$ Types 1, 2 and 3 respectively.

\begin{figure*}[htb]
 \includegraphics[width=0.32\linewidth]{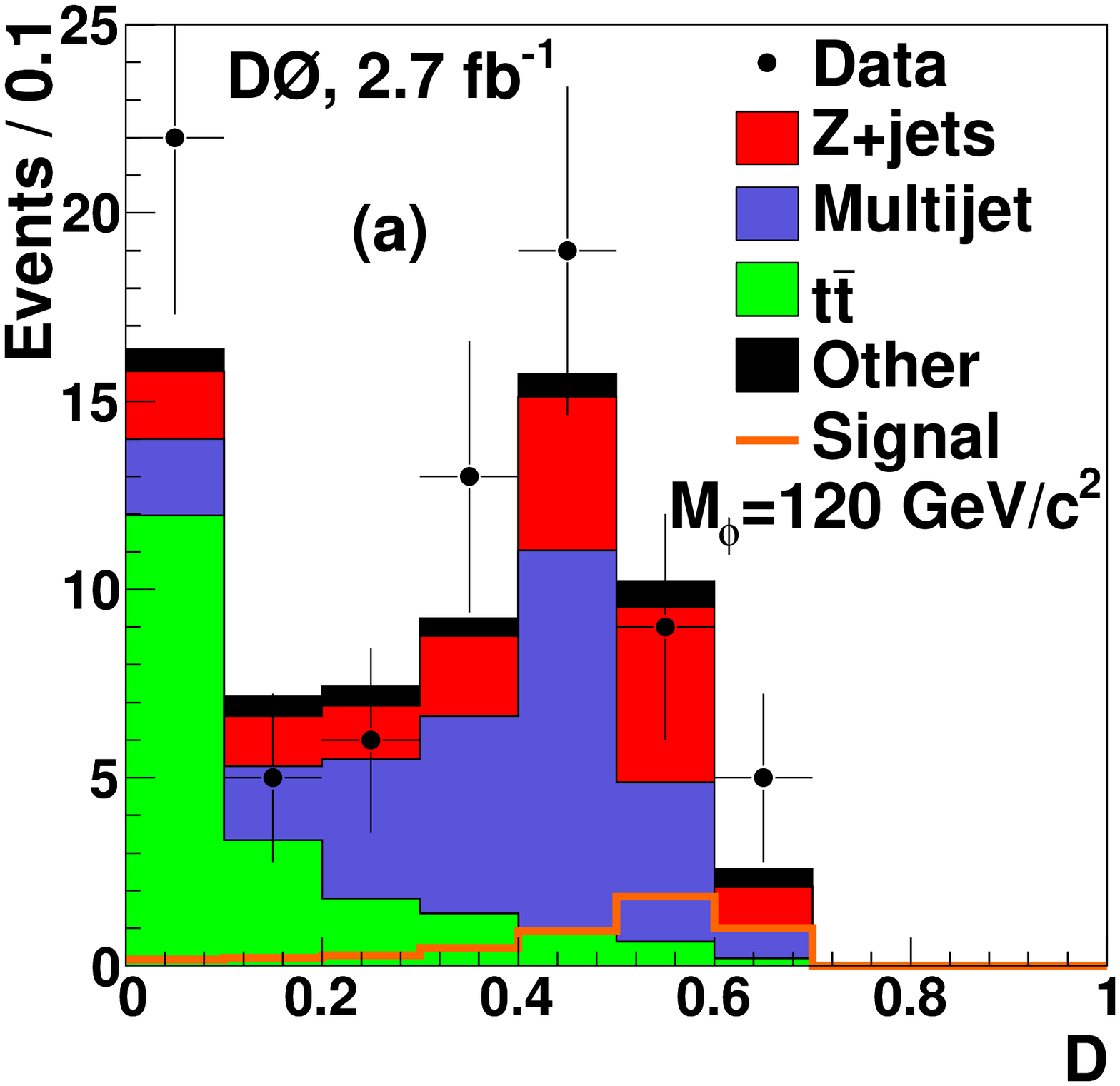}
 \includegraphics[width=0.32\linewidth]{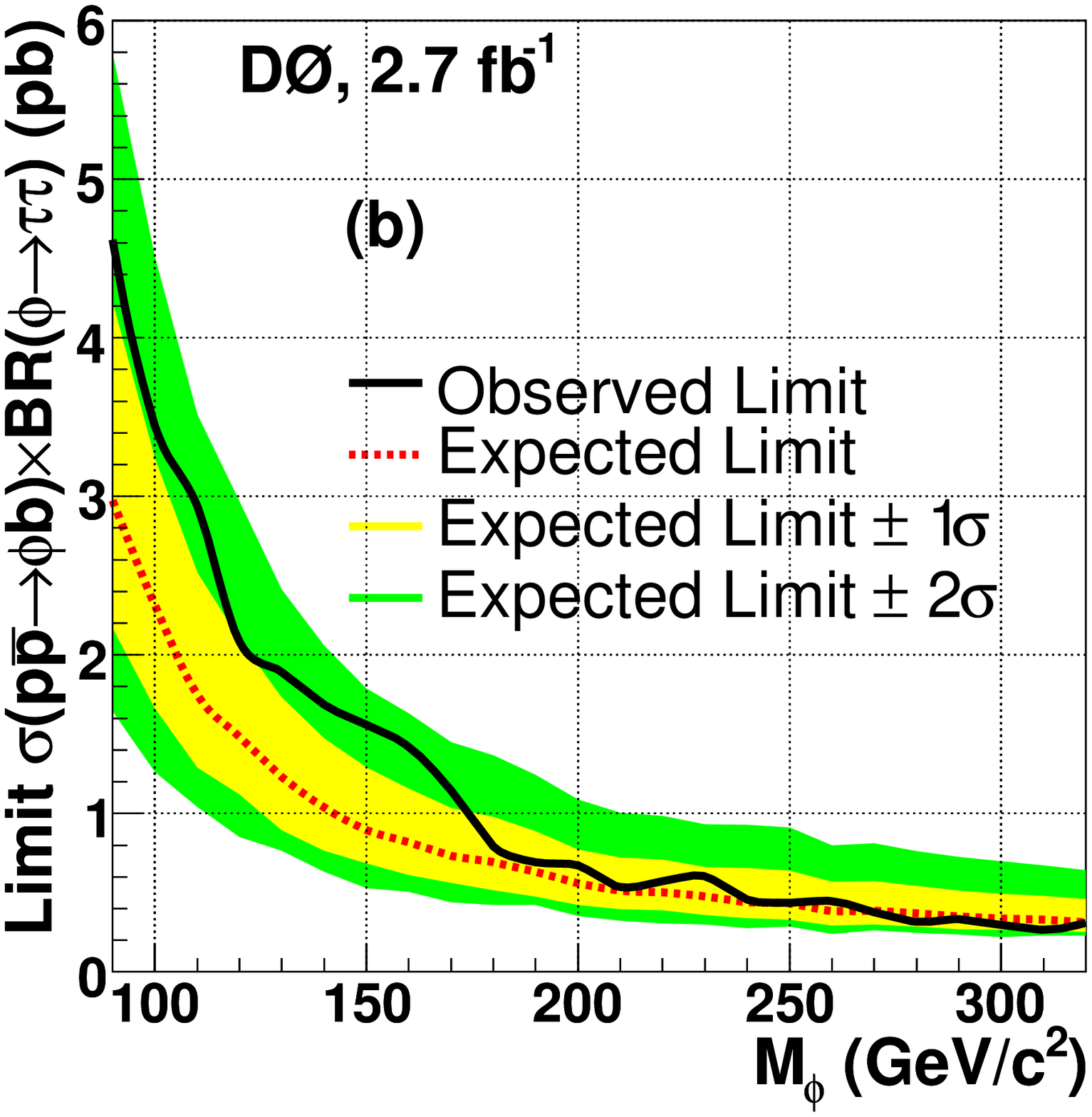}
 \includegraphics[width=0.32\linewidth]{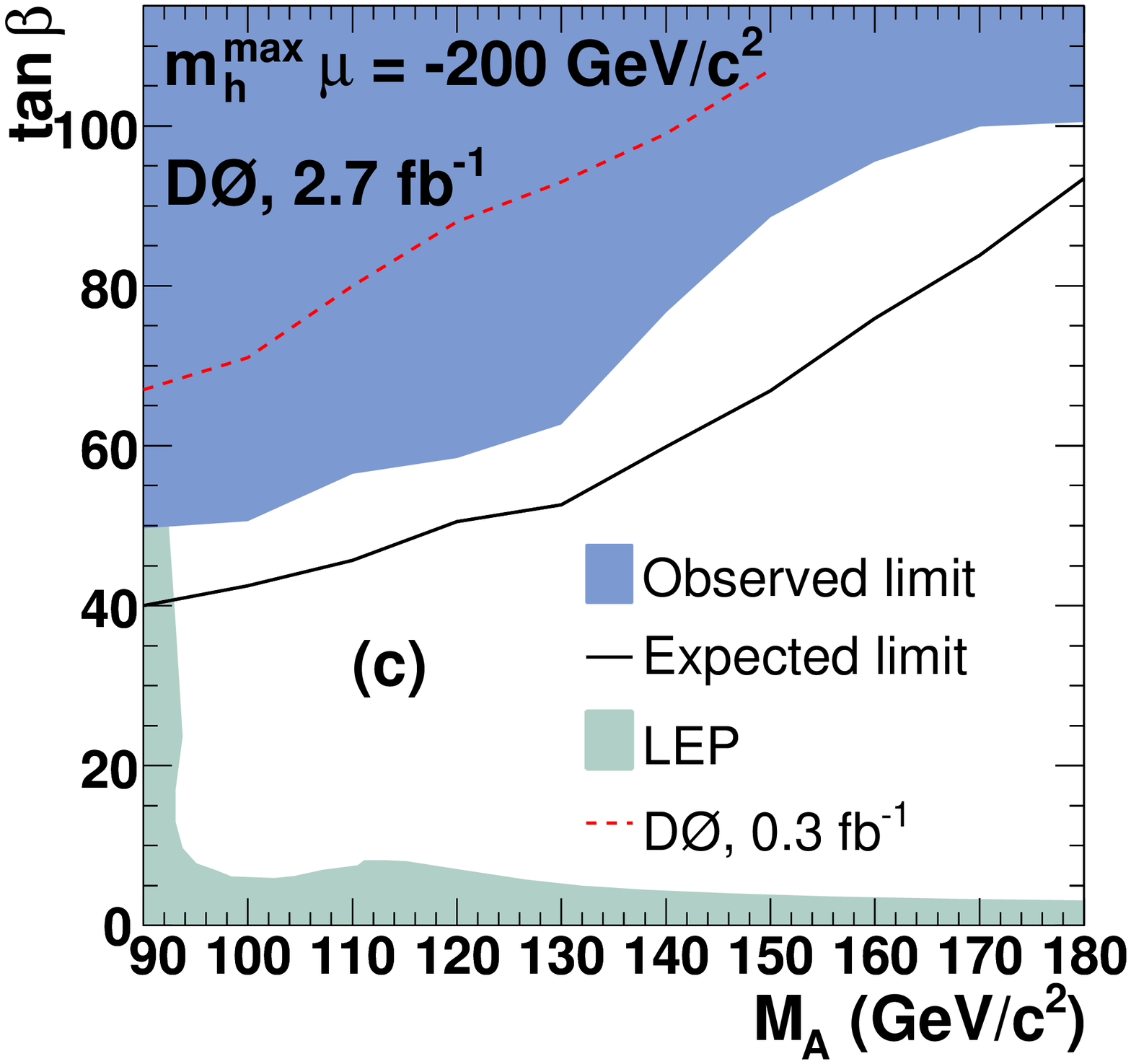}
 \caption{(a) The distribution of the final discriminant variable, $D=\mbox{ \em NN}_{top}\times\mbox{\em LL}_{MJ}$.  The figure 
          includes all $\tau$ Types.\label{f-prod-final}
    (b) The cross-section limit as a function of Higgs boson mass.\label{f-xsec}
    (c) The region in the $\tan\beta$ versus $M_A$ plane excluded by this analysis, LEP neutral MSSM Higgs searches, and the previous D0 result in this channel.\label{f-plane}}
\end{figure*}

After making the final selection, the discriminant $D$ is formed from the
product of the $\mbox{\em NN}_{top}$ and $\mbox{\em LL}_{MJ}$ variables, $D=\mbox{\em LL}_{MJ} \times \mbox{\em NN}_{top}$.  The resulting 
distributions for the predicted background, signal and
data are shown in Fig.~\ref{f-prod-final}(a).  This distribution is used as input
to a significance calculation using a modified frequentist approach with a
Poisson log-likelihood ratio test statistic~\cite{b-collie}.
In the absence of a significant signal we set 95\% confidence level limits on the presence of neutral 
Higgs bosons in our data sample.  The
cross section limits are shown in Fig.~\ref{f-xsec}(b) as a function of Higgs
boson mass.  These are translated into the $\tan\beta$ versus $M_A$ plane 
in the $m_h^{max}, \mu=-200$ GeV/$c^2$ MSSM benchmark scenario \cite{b-benchmark},
giving the excluded region shown in Fig.~\ref{f-plane}(c).  The signal cross sections and 
branching fractions are computed using \feynhiggs~\cite{b-sig-xsec}.
Instabilities in the theoretical calculation for $\tan\beta > 100$ limit the
usable mass range in the translation into the $(\tan\beta,\,M_A)$ plane.

In summary, this Letter reports a search for production of Higgs bosons in
association with a $b$ quark using eight times more data than previous results
for this channel.  The data are consistent with predictions from known physics
sources and limits are set on the neutral Higgs boson associated production cross section.  
These cross section limits, a factor of three improvement over previous results, 
are also translated into limits in the SUSY parameter space.


\input acknowledgement_paragraph_r2.tex   

\end{document}

%% file: list_of_authors_r2.tex
%
\author{V.M.~Abazov$^{37}$}
\author{B.~Abbott$^{75}$}
\author{M.~Abolins$^{65}$}
\author{B.S.~Acharya$^{30}$}
\author{M.~Adams$^{51}$}
\author{T.~Adams$^{49}$}
\author{E.~Aguilo$^{6}$}
\author{M.~Ahsan$^{59}$}
\author{G.D.~Alexeev$^{37}$}
\author{G.~Alkhazov$^{41}$}
% ...

\affiliation{\vspace{0.1 in}(The D\O\ Collaboration)\vspace{0.1 in}}
% ...
\affiliation{$^{6}$University of Alberta, Edmonton, Alberta, Canada;
                Simon Fraser University, Burnaby, British Columbia, Canada;
                York University, Toronto, Ontario, Canada and
                McGill University, Montreal, Quebec, Canada}
% ...
\affiliation{$^{30}$Tata Institute of Fundamental Research, Mumbai, India}
% ...
\affiliation{$^{37}$Joint Institute for Nuclear Research, Dubna, Russia}
% ...
\affiliation{$^{41}$Petersburg Nuclear Physics Institute,
                St. Petersburg, Russia}
% ...
\affiliation{$^{49}$Florida State University, Tallahassee, Florida 32306, USA}
% ...
\affiliation{$^{51}$University of Illinois at Chicago,
                Chicago, Illinois 60607, USA}
% ...
\affiliation{$^{59}$Kansas State University, Manhattan, Kansas 66506, USA}
% ...
\affiliation{$^{65}$Michigan State University,
                East Lansing, Michigan 48824, USA}
% ...
\affiliation{$^{75}$University of Oklahoma, Norman, Oklahoma 73019, USA}
% ...

%% file: acknowledgement_paragraph_r2.tex
%
We thank the staffs at Fermilab and collaborating institutions, 
and acknowledge support from the 
DOE and NSF (USA);
CEA and CNRS/IN2P3 (France);
FASI, Rosatom and RFBR (Russia);
CNPq, FAPERJ, FAPESP and FUNDUNESP (Brazil);
DAE and DST (India);
Colciencias (Colombia);
CONACyT (Mexico);
KRF and KOSEF (Korea);
CONICET and UBACyT (Argentina);
FOM (The Netherlands);
STFC and the Royal Society (United Kingdom);
MSMT and GACR (Czech Republic);
CRC Program, CFI, NSERC and WestGrid Project (Canada);
BMBF and DFG (Germany);
SFI (Ireland);
The Swedish Research Council (Sweden);
and
CAS and CNSF (China).